\documentstyle[aps,prl]{revtex}

\begin{document}
\title{\bf  Quantum Logic with a Single Trapped Electron}

\author{Stefano Mancini$^1$, Ana M. Martins$^2$ 
and Paolo Tombesi$^3$}

\address{
${}^1$Dipartimento di Fisica
and Unit\`a INFM,\\
Universit\`a di Milano, Via Celoria 16, 
I-20133 Milano, Italy\\
${}^2$Centro de Fisica de Plasmas, Instituto Superior Tecnico,\\ 
P-1096 Lisboa Codex, Portugal\\
${}^3$Dipartimento di Matematica e Fisica
and Unit\`a INFM, \\
Universit\`a di Camerino, 
I-62032 Camerino, Italy}

\date{Received: \today}

\maketitle

\begin{abstract}
We propose the use of a trapped electron to implement quantum logic 
operations.
The fundamental controlled-NOT gate is shown feasible.
The two quantum bits are stored in the internal and external 
(motional) degrees 
of freedom.
\end{abstract}


\widetext

\section{Introduction}
The modern theory of information relies on the very 
foundations of quantum mechanics. This is
because of information is physical, as recently 
emphasised by Landauer \cite{lan}. It implies
that the laws of quantum mechanics can be used 
to process and store information. The elementary
quantity of classical information is the bit, 
which is represented by a dichotomic system;
therefore, any physical realization of a bit 
needs a system with two states. The very novel
characteristics of quantum information is that, by using
quantum states to store information,
a quantum system can be in a superposition of
states. This means, in a sense, that the elementary
quantity of quantum information, a quantum bit, 
can be in both the states at the same time.

Already in 1981 Feynman \cite{feyn} pointed out the 
impossibility for a classical computer to
simulate the evolution of a quantum system in an 
efficient way. This opened the search of a more
efficient way to simulate quantum systems until
 Deutsch \cite{deu} provided a
satisfactory theoretical description of a universal 
quantum  computer. The quantum computer is a
device which operates with simple quantum logic gates. 
These are analogous to the
classical gates, which perform one elementary operation 
on two bits in a given way. Quantum logic
gates differ from their classical counterpart in that 
they operate on quantum superpositions and
perform operations on them \cite{divSci}.
It has also been shown that any quantum 
computation can be built from a series
of one-bit and two-bit quantum logic gates \cite{divPRA}. 
The fundamental quantum logic
gate is the controlled-NOT  (CN) gate \cite{fey,bar}, 
in which one quantum bit (or qubit) is flipped (rotated by
$\pi$ radians) depending upon the state of a second qubit. 

A very promising candidate for quantum logic 
was recently introduced by  
Cirac and Zoller \cite{cz}, who
showed how to construct universal multibit quantum logic gates 
in a system of laser-cooled
trapped ions. Other systems were devised as building blocks
for a quantum computer \cite{deueke}, the search for new systems is, 
however, still open because none of the previous systems is yet
claimed as the best candidate. One
should devise a system with very low loss, almost de-coherence free,
 which can be well controlled
with simple operations. However, before obtaining a suitable system one 
has to be sure that the mathematical models of quantum logic could
be easily implemented in a real physical system. Up to now the experimental realization 
of such logic operations were shown to be possible with trapped ions \cite{tur}, flying qubits
\cite{kimble}, and cavity QED \cite{dom}. There are claims that the quantum logic
gates are obtained in NMR systems \cite{cory} but this was also questioned \cite{caves}. In these 
systems, however, the implementation of quantum logic is not at all easy and was not completely 
performed in all of them.

It is here our aim to show that other natural candidates 
to implement quantum logic 
could be trapped electrons.
In fact, an electron is a real two-state system and when stored 
in a Penning trap \cite{penni} 
permits very
accurate measurements
\cite{van}. Furthermore, in such a system the decoherence effects, 
which can destroy the quantum interference that enables the 
quantum logic implementation \cite{deco}, are well
controlled \cite{bg}.
Moreover, electrons being structureless, 
open other possibilities, e.g. the use of statistics that has not as yet
been considered in the literature.

To introduce the system, in this paper we consider a single electron trapped in a 
Penning trap, and
we show how to get a controlled-NOT gate on a pair of qubits.
 The two qubits comprise two
internal (spin) states and two external (quantized
harmonic motion) states.
Although this minimal system consists of only two qubits, 
it illustrates the basic operations
necessary for, and the problems associated with, 
quantum logic networks with
electrons.
The extension to two o more electrons needs more investigations.
Here we are not interested in the scalability of the system, rather to show 
the physical implementation of quantum logic in a readly controllable way
with the existing technologies.

\section{The Model}

We are considering the
``geonium" system \cite{bg} consisting of an electron of 
charge $e$ and 
mass $m$ moving in 
a uniform 
magnetic field ${\bf B}$, along the positive $z$ axis, 
and a static 
quadrupole potential
\begin{equation}\label{V}
V=V_0\frac{x^2+y^2-2z^2}{4d^2}\,,
\end{equation}
where $d$ characterizes the dimension of the trap and $V_0$ 
is the potential applied to the trap
electrodes \cite{bg}.

In this work, in addition to the usual trapping fields, 
we embed the trapped electron
in a radiation field of vector potential ${\bf A}_{\rm ext}$.
Traditional hyperbolic Penning traps form
cavities for which it 
has not yet been possible to even
classify the standing-wave fields. In marked contrast, the 
radiation modes of a simple cylindrical
cavity are classified in a familiar way as either transverse 
magnetic or transverse electric modes
\cite{jac,gt}. So, in the following, we always refer to such 
cylindrical traps.

The Hamiltonian for
the trapped electron can be written as the 
quantum counterpart
of the classical Hamiltonian with the addition of the spin term
\begin{equation}\label{Hinit}
H=\frac{1}{2m}\left[{\bf p}
-e{\bf A}\right]^2
+eV-\frac{g}{2}\frac{e\hbar}{2m}\, \sigma \cdot{\bf B}\,,
\end{equation}
where $g$ is the electron's $g$ factor, and
\begin{equation}\label{A}
{\bf A}=\frac{1}{2}{\bf B}\wedge{\bf r}
+{\bf A}_{\rm ext}\,,
\end{equation}
where ${\bf r}\equiv(x,y,z)$, 
${\bf p}\equiv(p_x,p_y,p_z)$ 
are respectively the position and the
conjugate momentum operators, while 
$\sigma\equiv(\sigma_x,\sigma_y,\sigma_z)$
are the Pauli matrices in the spin space.

The motion of the electron in absence of the external field 
${\bf A}_{\rm ext}$ 
is the result of the motion of three
harmonic oscillators \cite{bg}, the cyclotron, the axial and 
the magnetron, well separated in the
energy scale, plus a spin precession around the $z$ axis. 
This can be 
easily understood by
introducing the ladder operators
\begin{eqnarray}\label{lad}
a_z&=&\sqrt{\frac{m\omega_z}{2\hbar}}\,z
+i\,\sqrt{\frac{1}{2\hbar m\omega_z}}\,p_z\\
a_c&=&\frac{1}{2}\left[\sqrt{\frac{m\omega_c}{2\hbar}}
(x-iy)
+\sqrt{\frac{2}{\hbar m\omega_c}}
(p_y+ip_x)\right]\\
a_m&=&\frac{1}{2}\left[\sqrt{\frac{m\omega_c}{2\hbar}}
(x+iy)
-\sqrt{\frac{2}{\hbar m\omega_c}}
(p_y-ip_x)\right]
\end{eqnarray}
where the indexes $z$, $c$ and $m$ stand for axial, cyclotron
and magnetron respectively.
The above operators obey the commutation relation 
$[a_{i},a^{\dag}_{j}]=\delta_{ij}$, $i, j=z,\,c,\,m$.

When ${\bf A}_{\rm ext}=0$, the Hamiltonian 
(\ref{Hinit}) simply reduces to
\begin{equation}\label{Hfree}
H=\hbar\omega_z a_z^{\dag} a_z
+\hbar\omega_c a_c^{\dag} a_c
-\hbar\omega_m a_m^{\dag} a_m
+\frac{\hbar}{2}\omega_s\sigma_z\,,
\end{equation}
where the angular frequencies are given by
\begin{equation}\label{freq}
\omega_z=\sqrt{\frac{|e|V_0}{md^2}}\,;\quad
\omega_c=\frac{|e|B}{m}\,;\quad
\omega_m\approx\frac{\omega_z^2}{2\omega_c}\,.
\end{equation}
and $\omega_s=g|e|B/2m$ is the spin precession angular frequency. 
In the previous expression for $\omega_c$ we neglected very small 
corrections \cite{bg}
which are not relevant for our purpose.
In typical experimental configurations \cite{bg} the respective 
frequency ranges are 
$\omega_z/2\pi \simeq$ MHz, $\omega_c/2\pi \simeq$ GHz, and
$\omega_m/2\pi \simeq$ kHz.

Let us introduce the external radiation field as a
standing wave along 
the $z$ direction 
and rotating, i.e. circularly polarized, in the $x-y$ plane
with frequency $\Omega$ \cite{mmt}.
In particular, we consider
a standing wave within the cilindrical 
cavity with wave vector $k$ 
and amplitude $|\alpha|$. Then, 
we can write 
\begin{equation}\label{Aext}
{\bf A}_{\rm ext}=\Bigg(
i\left[e^{i\varphi+i\Omega t}-e^{-i\varphi-i\Omega t}\right],
\left[e^{i\varphi+i\Omega t}+e^{-i\varphi-i\Omega t}\right],
0\Bigg)
\times |\alpha|\cos(kz+\phi)\,,
\end{equation}
where $\varphi$ is the phase of the wave field
which gives the direction of the electric (or magnetic)
vector in the $x-y$ plane at the initial time. We assume this can be 
experimentally controlled.
The amplitude $|\alpha|$ should
depend upon the transverse spatial variables 
through the Bessel function \cite{jac} but we can consider
it as a constant because of the small radius of 
the ciclotron motion\cite{gt}. The phase $\phi$
 definines the position of the 
center of the axial motion with respect
to the wave. Depending on its value the electron can
be positioned in any place between a node
and an antinode.

For frequencies $\Omega$ close to 
$\omega_c$ and $\omega_s$,  we can neglect the slow
magnetron motion, then the Hamiltonian (\ref{Hinit})
becomes
\begin{eqnarray}\label{Hnodip}
H&=&\hbar\omega_z a_z^{\dag} a_z
+\hbar\omega_c a_c^{\dag} a_c
+\frac{\hbar}{2}\omega_s\sigma_z\nonumber\\
&+&\hbar\epsilon\left[a_c e^{i\varphi+i\Omega t}
+a_c^{\dag}e^{-i\varphi-i\Omega t}\right]
\cos(k{\hat z}+\phi)\nonumber\\
&+&\hbar\zeta\left[
\sigma_-e^{i\varphi+i\Omega t}
+\sigma_+e^{-i\varphi-i\Omega t}
\right]
\sin(k{\hat z}+\phi)\,,
\end{eqnarray}
where
\begin{equation}\label{epze}
\epsilon
=\left(\frac{2|e|^3B}{\hbar m^2}\right)^{1/2}
|\alpha|\,,\quad 
\zeta=\frac{g|e|}{2m}|\alpha|k\,,
\end{equation}
and $\sigma_{\pm}=(\sigma_x\pm i\sigma_y)/2$.
The fourth and fifth terms in the right hand side of the 
Hamiltonian
(\ref{Hnodip}) describe the interaction between 
the trapped electron and the standing wave
which can give rise to a coupling between the axial 
and cyclotron
motions, as well as between the axial and spin ones.
In writing Eq. (\ref{Hnodip}) we omitted terms coming from 
${\bf A}_{\rm ext}^2$ which give a negligible contribution
(at most an axial frequency correction) 
when the electron in positioned in a node or antinode as we shall
do in the following.

\section{Entangled States Preparation}

The spin state is usually controlled through a small 
oscillatory magnetic 
field ${\bf b}$ that lies in the
$x-y$ plane \cite{bg}
\begin{equation}\label{bfield}
{\bf b}(t)=b\Big(\cos(\omega_s t+\theta),\,\sin(\omega_s t+\theta),
\,0\Big)\,,
\end{equation}
which causes Rabi oscillations at frequency $\varpi_s=g|e|b/2m$. The 
phase $\theta$ can be experimentally
controlled; it gives the direction of the field at initial times .
The Hamiltonian that follows
from Eq. (\ref{bfield}), in  absence of the standing wave and 
in a frame 
rotating at frequency $\omega_s$, is
\begin{equation}\label{Hb}
H_s=\hbar\frac{\varpi_s}{2}\left[\sigma_+ e^{-i\theta}
+\sigma_- e^{i\theta}\right]
=\hbar\frac{\varpi_s}{2}\left[\sigma_x\cos\theta
+\sigma_y\sin\theta\right]\,.
\end{equation}
The other non interacting terms do not affect the spin motion
and can be neglected.
The evolution of the spin state $|\chi\rangle_s=u|\uparrow\rangle
+v|\downarrow\rangle$, with $|u|^2+|v|^2=1$, 
under such Hamiltonian will be
\begin{equation}\label{chit}
|\chi(t)\rangle_s=
\left[u\cos\left(\frac{\varpi_s t}{2}\right)
-ive^{-i\theta}\sin\left(\frac{\varpi_s t}{2}\right)\right]|
\uparrow\rangle
+\left[v\cos\left(\frac{\varpi_s t}{2}\right)
-iue^{i\theta}\sin\left(\frac{\varpi_s t}{2}\right)\right]|
\downarrow\rangle\,.
\end{equation}
Thus, depending on the interaction time, any superposition of spin
states can be generated.

For what concerns the spatial degrees of freedom, we assume  
 the cyclotron and the axial motions are deep cooled down
 to their respective lower states, i.e. $|0\rangle_c$ 
and $|0\rangle_z$. This could be achievable when 
the axial motion is decoupled from  the external 
circuit usually used 
to extract information \cite{bg,gt}.

We now consider the spin and the axial degrees of freedom as qubits.
Then, by choosing $\phi=0$, i.e. positioning the electron 
in the node
of the standing wave,
 Eq. (\ref{Hnodip}) can be approximated by
\begin{eqnarray}\label{Happrox}
H&=&\hbar\omega_z a_z^{\dag} a_z
+\hbar\omega_c a_c^{\dag} a_c
+\frac{\hbar}{2}\omega_s\sigma_z\nonumber\\
&+&\hbar\epsilon\left[a_c e^{i\varphi+i\Omega t}
+a_c^{\dag}e^{-i\varphi-i\Omega t}\right]
\nonumber\\
&+&\hbar\zeta k \sqrt{\frac{\hbar}{2m\omega_z}}
\left[
\sigma_-e^{i\varphi+i\Omega t}
+\sigma_+e^{-i\varphi-i\Omega t}
\right]
\left(a_z+a_z^{\dag}\right)\,.
\end{eqnarray}

We distinguish two situations (in a frame rotating
at frequency $\Omega$): the first one in which $\Omega
=\omega_s-\omega_z$ gives

\begin{equation}\label{Hint1}
H_-=\hbar\eta\left[\sigma_+a_ze^{-i\varphi}
+\sigma_-a_z^{\dag}e^{i\varphi}\right]\,,
\end{equation}
where $\eta=k\zeta\sqrt{\hbar/2m\omega_z}$.

The second, for which $\Omega=\omega_s+\omega_z$ gives
\begin{equation}\label{Hint2}
H_+=\hbar\eta\left[\sigma_+a_z^{\dag}e^{-i\varphi}
+\sigma_-a_ze^{i\varphi}\right]\,.
\end{equation}

The action of Hamiltonian (\ref{Hint1}) for a time $t$ over 
an initial state
$|0\rangle_z|\uparrow\rangle$ leads to
\begin{equation}\label{evHint1}
|0\rangle_z|\uparrow\rangle\to
\cos(\eta t)|0\rangle_z|\uparrow\rangle
-ie^{i\varphi}\sin(\eta t)|1\rangle_z|\downarrow\rangle\,.
\end{equation}

Instead, the action of Hamiltonian (\ref{Hint2}) for a time 
$t$ over an initial state
$|0\rangle_z|\downarrow\rangle$ leads to
\begin{equation}\label{evHint2}
|0\rangle_z|\downarrow\rangle\to
\cos(\eta t)|0\rangle_z|\downarrow\rangle
-ie^{-i\varphi}\sin(\eta t)|1\rangle_z|\uparrow\rangle\,.
\end{equation}

Practically, if the electron enters in the trap with e.g. its spin 
down, 
by applying selectively the Hamiltonians (\ref{Hb}), 
(\ref{Hint1}) and (\ref{Hint2}) for appropriate times
we can get states of the form
\begin{equation}\label{stategen}
\alpha|0\rangle_z\,|\downarrow\rangle
+\beta|0\rangle_z\,|\uparrow\rangle
+\gamma|1\rangle_z\,|\downarrow\rangle
+\delta|1\rangle_z\,|\uparrow\rangle\,,\quad
|\alpha|^2+|\beta|^2+|\gamma|^2+|\delta|^2=1\,,
\end{equation}
which show entanglement between the two qubits.

Therefore, the manipulation between the 
four basis eigenstates spanning the
two-qubit register
${\cal B}\equiv\{|0\rangle_z\,|\downarrow\rangle\,,\;
|0\rangle_z\,|\uparrow\rangle\,,\;
|1\rangle_z\,|\downarrow\rangle\,,\;
|1\rangle_z\,|\uparrow\rangle\,\}$
is achievable.

\section{Logic Operations}

Here we shall consider the spin as ``target" qubit, and the
axial degree as ``control" qubit.
The basic logic operations on a single qubit (e.g. Hadamard gate)
can be implemented in the target qubit by applying the Hamiltonian
(\ref{Hb}), while there is no way to control directly the 
axial qubit.

The CN gate represents, instead, a computation at the most
fundamental  level: the target qubit is flipped
depending upon the state of the control qubit. 

The truth table of the reduced CN gate is
\begin{eqnarray}\label{CNtable}
|0\rangle_z\,|\downarrow\rangle &\to& |0\rangle_z\,
|\downarrow\rangle\,,\nonumber\\
|0\rangle_z\,|\uparrow\rangle &\to& |0\rangle_z\,
|\uparrow\rangle\,,\nonumber\\
|1\rangle_z\,|\downarrow\rangle &\to& |1\rangle_z\,
|\uparrow\rangle\,,\nonumber\\
|1\rangle_z\,|\uparrow\rangle &\to& |1\rangle_z\,
|\downarrow\rangle\,.
\end{eqnarray}

To implement such a transformation we consider $\Omega=\omega_s$
and $\phi=-\pi/2$, i.e. the electron is positioned in an antinode 
(this operation is routinely performed in actual 
experiments \cite{gt}).
Then, the leading term of Eq.
(\ref{Hnodip}) (in a frame rotating at frequency $\Omega$) 
will result
\begin{equation}\label{HintCN1}
H=-\hbar\zeta
\left[\sigma_+e^{-i\varphi}
+\sigma_-e^{i\varphi}\right]
\times\left[1-\frac{\hbar k^2}{4m\omega_z}
-\frac{\hbar k^2}{2m\omega_z}a_z^{\dag}a_z
\right]\,.
\end{equation}

If we choose $\varphi=0$, the above Hamiltonian reduces to
\begin{equation}\label{HintCN2}
H=-\hbar 2\zeta\left(
1-\frac{\hbar k^2}{4m\omega_z}\right)\sigma_x
+\hbar 2\zeta\frac{\hbar k^2}{2m\omega_z}a_z^{\dag}a_z\sigma_x\,.
\end{equation}

Of course, for logic operations on the two qubits, only the 
interacting part of the above
Hamiltonian is relevant.
On the other hand the flipping effect
of the first term of
Hamiltonian (\ref{HintCN2}) 
can be eliminated by a successive action of Hamiltonian (\ref{Hb}) 
with $\theta=0$, for a time $\tau$ such that
\begin{equation}\label{cond}
\tau \varpi_s=4\zeta\left(1-\frac{\hbar k^2}{4m\omega_z}\right)
t^*\pm 2\pi n\,,
\end{equation}
where $n$ is a natural number and $t^*$ is the interaction time 
with Hamiltonian 
(\ref{HintCN2}).

Hence, the relevant Hamiltonian for the CN gate is
\begin{equation}\label{Heff}
H=\hbar\kappa \, a^{\dag}_z a_z\, \sigma_x\,,
\end{equation}
where $\kappa=\hbar\zeta k^2/m\omega_z$.

If we appropriately choose the interaction time 
$t^*=\pi/2\kappa$ we can 
apply the transformation 
\begin{equation}\label{UCN}
U=\exp\left(-i\pi\, a^{\dag}_z a_z\, \sigma_x/2\right)\,.
\end{equation}
Thus, the net unitary transformation, in the 
${\cal B}$ basis, is
\begin{eqnarray}\label{CNtra}
\left(
\begin{array}{cccc}
1,0,0,0\\
0,1,0,0\\
0,0,0,-i\\
0,0,-i,0
\end{array}
\right)\,.
\end{eqnarray}

This transformation is equivalent to the reduced 
CN gate of Eq. (\ref{CNtable}), apart from phase
factors that can be eliminated by the appropriate 
phase settings of subsequent logic operations
\cite{bar}. Practically, the reduced 
CN gate consists here in a single step similarly to 
Ref. \cite{monRC}.

\section{Information Measurements}

We recall that in the geonium system the 
measurements are performed on the axial 
degree of freedom due to the nonexistence of good detectors 
in the microwave regime  \cite{bg}. The oscillating charged
particle induces alternating image charges on the electrodes, 
which in turn cause an oscillating
current to flow through an external circuit where the measurement
is performed. 
The current will 
be proportional to the axial momentum
$p_z$ \cite{bg}. 
The very act of measurement changes, however, the
state of the measured observable. Then, in order not to
loose any stored information because of the measurement,
we shall transfer 
the information contained in the axial qubit  into the 
cyclotron degree of freedom
prior to the measurement procedure. This will allow us to get a 
complete information about 
the qubits by coupling different cyclotron and spin 
observables with the axial degree of freedom.

To transfer the information from the axial motion to the cyclotron 
one, we again use the standing wave,
but with another resonance, $\Omega=\omega_c-\omega_z$ in 
order to get from Eq. (\ref{Hnodip})
\begin{equation}\label{Htransf}
H=i\hbar\epsilon k\sqrt{\frac{\hbar}{2m\omega_z}}
\left(a_c^{\dag}a_z-a_ca_z^{\dag}\right)\,.
\end{equation}
Here we set $\phi=\varphi=-\pi/2$.
With the action of the Hamiltonian (\ref{Htransf}) for a well 
chosen interaction time, it is 
possible to transfer any previously entangled
state as follows
\begin{equation}\label{transf}
|0\rangle_c\Big[c_0|0\rangle_z|\chi\rangle_s
+c_1|1\rangle_z|\chi'\rangle_s\Big]
\to
\Big[c_0|0\rangle_c|\chi\rangle_s
+c_1|1\rangle_c|\chi'\rangle_s\Big]|0\rangle_z\,,
\end{equation}
where $|\chi\rangle$ and $|\chi'\rangle$ represent 
two generic spin states. 
This is obtained when the interaction time is 
$t=\sqrt{\pi m \omega_z/2\hbar\epsilon k}$.

Once the information is transferred to the cyclotron 
degree of freedom, the axial 
motion is coupled with the external
circuit, and it will reach the thermal equilibrium with the read-out 
apparatus.

Then, the measurements of $a^{\dag}_c a_c$ and $\sigma_z$ can be
done in the usaul way with the aid of the magnetic bottle 
which causes 
a shift of the axial resonance
proportional to the respective quantum numbers \cite{bg}
\begin{equation}\label{shift}
\Delta\omega_z\approx{\tilde\omega_z}
\left(\frac{gs}{4}+n_c+\frac{1}{2}\right)\,,
\end{equation}
where $\tilde\omega_z$ is a constant, and $n_c$, $s$ are 
the cyclotron excitacion and spin quantum numbers.
This frequency shift can be measured with very
high precision \cite{bg}.

In this model it could be also possible to obtain phase information 
about the quantum state of the register by means of 
the coupling between the meter (axial degree) 
and the system (cyclotron or spin) induced
again by the standing waves (see e.g. Ref.\cite{mmt}).

\section{Conclusions}

In conclusion, we have shown the possibility of using a trapped 
electron for fundamental quantum
logic. That system has the advantage of a well defined and simple 
internal structure and,
practically, the decoherence appears only in the axial degree of 
freedom as a consequence of
measurements but the information stored in this degree of freedom,
prior to the measurement,
can be unitarily transferred into the cyclotron motion.
The latter can be preserved from decoherence due to decay mechanisms 
by appropriately tuning the cavity \cite{gd}. 
The spin is very stable against fields fluctuations \cite{gdk}.
Eventually, the register ${\cal B}$, in such a configuration,  
could only suffer of the time uncertainty
in the switching on and off the interactions, possibly leading to 
nondissipative decoherence \cite{mil,boni}.
The effect on the fidelity in performing the logical operations 
could arise, indeed, from the impurity of the 
motional ground states 
due to an imperfect cooling process. 
Anyway, we retain that the present model can be  
implemented with the 
current technology, and a
comparison with the results obtained in the experiment of Ref. 
\cite{monPRL} would be useful. With respect to the last Reference
in the present case the complete information on the state 
of the two-qubit register is also obtainable. 

We also whish to remark that, within the model of trapped electron, 
other schemes could be exploited, 
for example by encoding information in other degrees, or by using Schroedinger
cat states as well
\cite{coc}; in fact the latter were shown to be  achievable in such 
systems \cite{mmt,ourRC}.

The next step would be the extension of the above formalism to 
the case of two or more trapped electrons,
in order to investigate real possibilities for 
quantum registers. One should consider that the realization of a 
4-qubit system would be a real advancement because of the possibility of
checking error correction strategies. As a final comment we can say 
that with this simple system we have introduced here, one can implement 
\cite{martom} the Deutsch problem \cite{deu,artur} as well.

\bigskip

The authors are grateful for a critical reading of the manuscript by I. Marzoli.
This work has been partially supported by INFM (through 
the 1997 Advanced Research Project ``CAT''), by the
European Union in the framework of the TMR Network ``Microlasers
and Cavity QED'', by MURST under the ``Cofinanziamento 1997'' and
by the CNR-ICCTI joint programme.

\bibliographystyle{unsrt}

\begin{thebibliography}{88}

\bibitem{lan} R. Landauer, Phys. Today {\bf 44}({\bf 5}), 23 (1991)

\bibitem{feyn} R. Feynman, Int. J. Theor. Phys.{\bf 21}, 467, (1982)

\bibitem{deu} 
D. Deutsch, Proc. R. Soc. London A {\bf 400}, 97, (1985), 
{\it ibid.} {\bf 425}, 73, (1989)

\bibitem{divSci}
D. P. DiVincenzo, Science {\bf 270}, 255 (1995);
S. Lloyd, Sci. Am. (Int. Ed.) {\bf 273}, 140 (1995);
A. Ekert and R. Jozsa, Rev. Mod. Phys. {\bf 68}, 733 (1995).

\bibitem{divPRA}
D. P. DiVincenzo, Phys. Rev. A {\bf 51}, 1015 (1995);
A. Barenco {\it et al., ibid.}, {\bf 52}, 3457 (1995);
S. Lloyd, Phys. Rev. Lett. {\bf 75}, 346 (1995).

\bibitem{fey}
R. P. Feynman, Opt. News {\bf 11}, 11 (1985).

\bibitem{bar}
A. Barenco, D. Deutsch, A. Eckert and R. Jozsa, Phys. Rev. Lett.
{\bf 74}, 4083 (1995).

\bibitem{cz}
J. I. Cirac and P. Zoller, Phys. Rev. Lett. {\bf 74}, 4091 (1995).

\bibitem{deueke}
D. Deutsch and A. Ekert,  Phys. World {\bf 11}, 47 (1998)

\bibitem{tur}
Q.A. Turchette, C.S. Wood, B.E. King, C.J. Myatt, D. Leibfried, W. M. Itano,
C. Monroe, and D. J. Wineland, Phys. Rev. Letts. {\bf 81}, 3631 (1998).  

\bibitem{kimble}
Q. A. Turchette, C. J. Hood, W. Lange, H. Mabuchi, and  H. J. Kimble,
Phys. Rev. Lett. {\bf 75}, 4710 (1995).

\bibitem{dom}
P. Domokos, J. M. Raimond, M. Brune and S. Haroche, Phys. Rev. A {\bf 52}, 
3554 (1995).

\bibitem{cory}
D. G. Cory {\it et al.} Phys. Rev. Lett. {\bf 81}, 2152 (1998).

\bibitem{caves}
R. Shack and C. M. Caves,
e-print: quant-ph/9903101.

\bibitem{penni}
F. M. Penning, Physica (Amsterdam) {\bf 3}, 873 (1936).

\bibitem{van}
R. S. Van Dyck, Jr.,  P. B. Schwinberg and H. Dehemelt,
in {\it Atomic Physics 9}, edited by R. S. Van Dyck, 
Jr. and E. N. Fortson
(World Scientific, Singapore, 1984);
R. S. Van Dyck, Jr.,  P. B. Schwinberg and H. Dehemelt,
Phys. Rev. Lett. {\bf 59}, 26 (1987);
R. S. Van Dyck, Jr., in {\it Quantum Electrodynamics}, edited by
T. Kinoshita (World Scientific, Singapore, 1990).

\bibitem{deco}
W. H. Zurek, Phys. Today {\bf 44}({\bf 10}), 36 (1991);
R. Landauer, in Proc. Drexel Fourth Symposium on 
Quantum Nonintegrability-Quantum Classical
Correspondence,  D. H. Feng and B. L. Hu Eds. 
(International Press, Boston, 1996);
W. G. Unruh,  Phys. Rev. A {\bf 51}, 992 (1995).

\bibitem{bg}
L. S. Brown and G. Gabrielse, Rev. Mod. Phys. {\bf 58}, 233 (1986).


\bibitem{jac} 
J. D. Jackson, {\it Classical Electrodynamics}, 
(Wiley, New York, 1975).

\bibitem{gt}
G. Gabrielse and J. Tan, in {\it Cavity Quantum Electrodynamics},
edited by P. R. Berman, (Academic Press, San Diego CA, 1994), p.267.

\bibitem{mmt}
A. M. Martins, S. Mancini and P. Tombesi, Phys. Rev. A {\bf 58}, 
3813 (1998). 

\bibitem{monRC}
C. Monroe, D. Leibfried, B. E. King, D. M. Meekhof, 
W. M. Itano and D. J. Wineland,
Phys. Rev. A {\bf 55}, R2489 (1997).

\bibitem{mt}
S. Mancini and P. Tombesi,
Phys. Rev. A {\bf 56}, 3060 (1997).

\bibitem{gd}
G. Gabrielse and H. G. Dehmelt,
Phys. Rev. Lett. {\bf 55}, 67 (1985).

\bibitem{gdk}
G. Gabrielse, H. G. Dehmelt and W. Kells, Phys. Rev. Lett.
{\bf 54}, 537 (1985).

\bibitem{mil}
G. J. Milburn, Phys. Rev A {\bf 44}, 5401 (1991).

\bibitem{boni}
R. Bonifacio, Nuovo Cimento B {\bf 114}, 473 (1999),
eprint: quant-ph/ 9901063.

\bibitem{monPRL}
C. Monroe, D. M. Meekhof, 
B. E. King, W. M. Itano and D. J. Wineland,
Phys. Rev. Lett. {\bf 75}, 4714 (1995).


\bibitem{coc}
P. T. Cochrane, G. J. Milburn and W. J. Munro,
eprint: quant-ph/9809037.


\bibitem{ourRC}
S. Mancini and P. Tombesi,
Phys. Rev. A {\bf 56}, R1679 (1997).

\bibitem{martom} 
I. Marzoli and P. Tombesi (unpublished).

\bibitem{artur}
A. Ekert and C. Macchiavello in {\it Unconventional Models
of Computation}, edited by C.S. Claude, J. Casti and M.J. Dinnen, pp. 19-44.
Springer Series in Discrete Mathematics and Theoretical Computer Science
(Springer, Singapore, 1998).




\end{thebibliography}

\end{document}